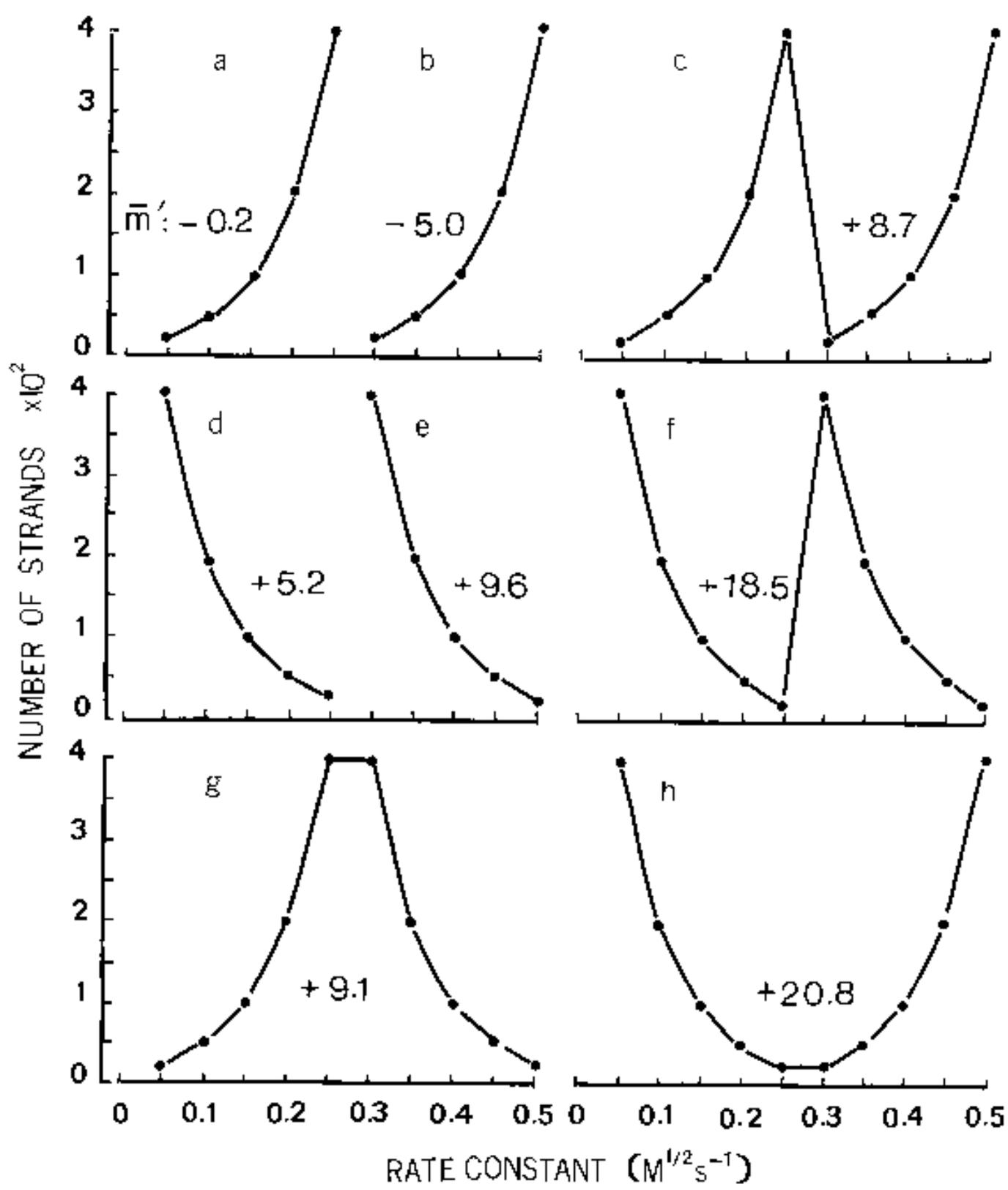

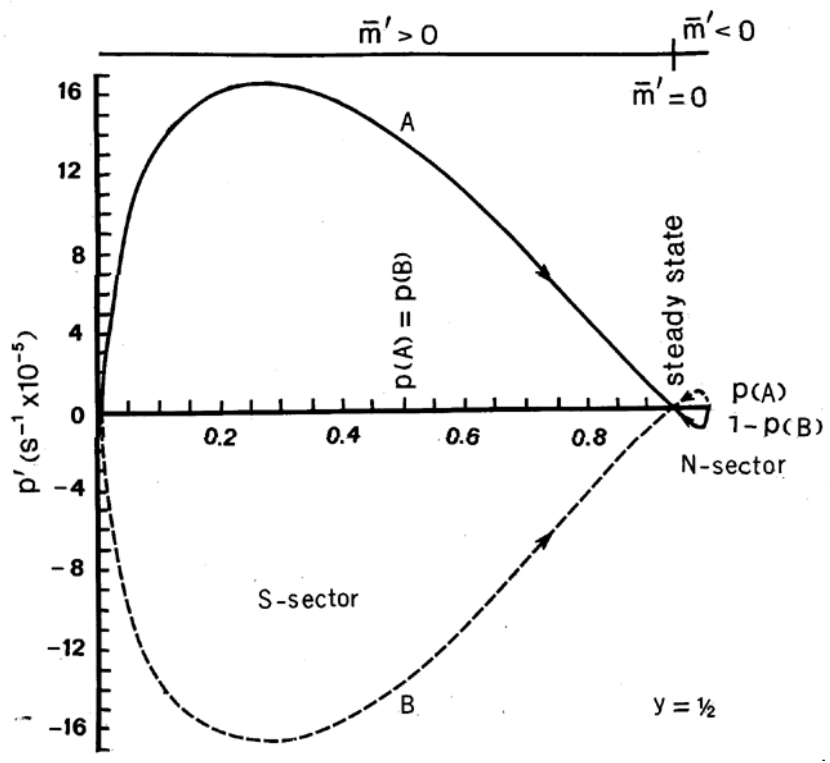

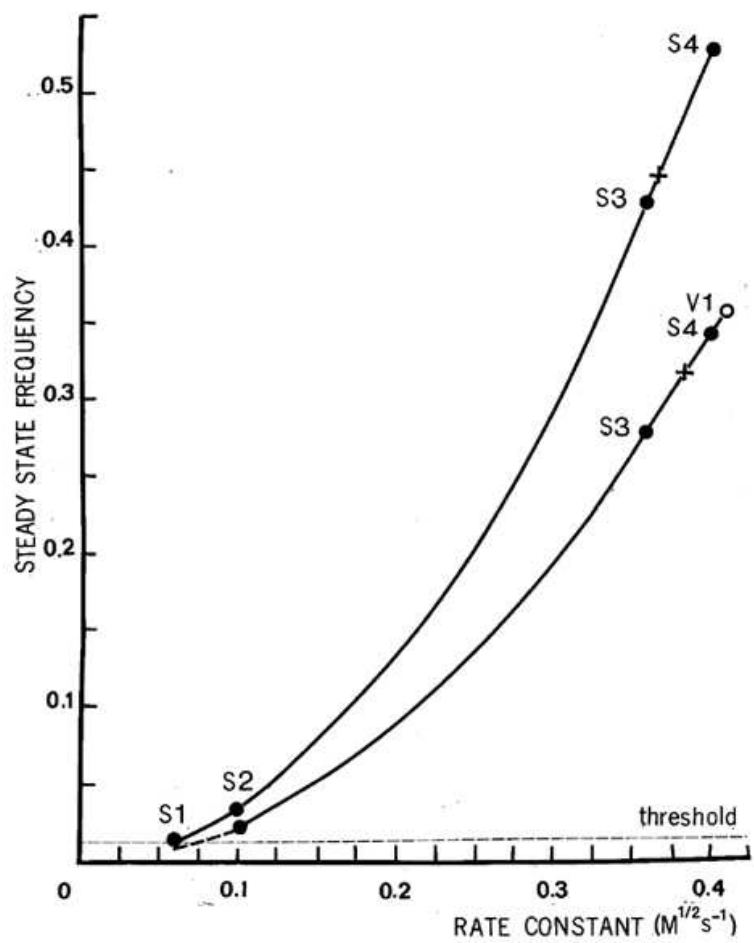

DARWINIAN ASPECTS OF MOLECULAR EVOLUTION

AT SUBLINEAR PROPAGATION RATES

Brian K. Davis

Research Foundation of Southern California, Inc.

5580 La Jolla Boulevard

La Jolla, CA 92037, U.S.A.

4 Figures

Summary

The symmetric distribution and all other states in the symmetry sector of the frequency trajectory increase mean fitness during competitive replication at sublinear propagation rates (parabolic time course). States in the non-symmetry sector, by contrast, produce negative time variations in mean fitness. The polymorphic steady state attained in sublinear systems is destabilised by formation of a variant with above threshold fitness. Evolution in the post-steady state interval increases threshold fitness. Contrary to the proposition that 'parabolic growth invariably results in the survival of all competing species,' only species with sufficient fitness to avoid subthreshold frequencies survive.

1. Introduction

Pre-replication, prebiotic and biological stages of evolution can be interpreted as a damping response to suitably scaled thermodynamic and kinetic (selection) forces (Davis, 1996a, 1998). In addition to including the early stages that preceded self-propagation, the statistical physics of evolution also provides a theoretical basis for both gradual and episodic transitions in the origin of species. Since pre-replication events and non-gradual transitions are alien to a theory of evolution based on comparative doubling rates among competing alleles or polymers, the proposed theory has wider scope.

Damping of the kinetic force during competitive replication was initially restricted to positive time variations in the mean formation rate constant (mean fitness). The validity of this condition was assured at linear rates of propagation (Fisher, 1930; Davis, 1978). At sub-linear propagation rates, mean fitness also increases, as required, when species frequencies are symmetrically distributed. Since sub-linear systems include non-symmetric states and can also decrease mean fitness over time, this quasi-linear theory, restricted to linear force damping relations at sub-linear kinetics, plainly represented an incomplete model. While force damping relations unconstrained by this requirement are now established (Davis, 1998), the finding that propagation at sub-linear rates resembles Darwinian evolution in certain aspects remains of interest.

Fractional order ($0 < y < 1$) kinetics occur in replicase-free, template-directed synthesis of DNA or RNA oligomers (von Kiedrowski, 1986; Zielinski and Orgel, 1987; von Kiedrowski et al., 1991). These processes are relevant to pre-replicase stages in the origin of life. Synthesis follows a parabolic time course, because the rate of second order annealing between free complementary strands accelerates with advancement of synthesis. As duplex molecules sre synthetically inactive, they increasingly retard synthesis. Szathmary and Gladkih (1989) found that competitive replication at sub-linear propagation rates evolves toward a non-Darwinian attractor. A polymorphic steady state was discovered to replace 'survival of the fittest'. Their condition for species coexistence was subsequently revised to remove an inaccuracy (Davis, 1996b). The concept of an attractor at non-limit frequencies ($p \neq 1, 0$), nonetheless, remained valid.

In a second departure from Darwinian systems, sub-linear systems were demonstrated to optimize the $y$-mean of fitness (excess production term), rather than arithmetic mean (Davis, 1996a, Appendix B). An extremum reported by Varga and Szathmary (1997) is a polymer-scale version of this principle. Holder's inequality was used to obtain this result, while the former study relied on variational calculus.

Despite these departures from Darwinian evolution, it will be shown that the symmetry-related increase in mean fitness extends to all states in the symmetry sector of the frequency trajectory during competitive replication at sub-linear propagation rates. In addition, it is demonstrated that the fitness threshold for species survival is non-decreasing in sub-linear systems. The apparent source of a misinterpretation (Varga and Szathmary, 1997) regarding a term in the equation for time variations in mean fitness is identified. Broadly, the proposition that molecular evolution at sub-linear propagation rates (parabolic growth) 'invariably results in the survival of all competing types' is disproved and this process is shown to be far more Darwinian than previously realized.

## 2. Mean fitness increases in symmetric frequency distributions

The equation for rate of change in mean fitness during competitive replication at sub-linear propagation rates was previously demonstrated to imply that time variations in mean fitness are positive, when species frequencies are symmetrically distributed (Davis, 1996a, p. 88). Figures 1a,b,c illustrate the power of this symmetry principle. Negative rates of change in mean fitness ($-0.2 \times 10$ and

FIGURE 1

$-5.0 \times 10^{-4}$ $M^{\frac{1}{2}}s^{-2}$) in two frequency distributions convert to a positive rate of change in mean fitness ($+8.7 \times 10^{-4}$ $M^{\frac{1}{2}}s^{-2}$), when both distributions are combined to form a distribution with translational symmetry. Counter-Darwinian time variations in mean fitness at ½-order kinetics are thus suppressed by symmetry.

Figures 1d,e show that inversion of the distributions in Figs. 1a,b yields two non-symmetric distributions with positive rates of change in mean fitness (+5.2x10$^{-4}$ and +9.6x10$^{-4}$ M$^{½}$s$^{-2}$). They also yield a positive rate of change in mean fitness, when combined into a symmetric distribution. The merged distribution displays a strongly positive rate of change in mean fitness (+18.5x10$^{-4}$ M$^{½}$s$^{-2}$). Symmetry can therefore produce positive time variations in mean fitness at fractional order kinetics, whether the initial frequency distribution exhibits a Darwinian or counter-Darwinian pattern of evolution.

As might be anticipated, positive time variations in mean fitness are exhibited by distributions with other forms of symmetry. In Figs. 1g,h, two distributions with mirror symmetry (concave-upright bell and convex-inverted bell) increase mean fitness at rates of +9.1x10$^{-4}$ and +20.8x10$^{-4}$ M$^{½}$s$^{-2}$.

### 3. Symmetry-sector of the frequency trajectory

When a system evolves toward a stable polymorphic steady state (Szathmary and Gladkih, 1989; Davis, 1996b, 1998; Varga and Szathmary, 1997), the direction of change in mean fitness with time depends on the initial state. One set of species frequencies will yield positive time variations in the mean, as a low initial fitness is elevated to its steady state value. Another set produces negative rates of change in mean fitness, lowering the mean to its steady state value. The former set of frequencies can be expected from Section 2 to include the symmetry distribution.

The trajectory of each frequency distribution during pairwise competitive replication at ½-order kinetics is depicted in Fig. 2. These curves give the rate of change in frequency at different species frequencies (Davis, 1998). They were obtained on comparing DNA synthesis in two replicase-free preparations in the study of Sievers and von Kiedrowski (1994). Template-directed synthesis of dCCGCCG (dAA) and its complementary hexamer, dCGGCGG (dBB), versus that of dCCGCGG (dAB) (Sievers and von Kiedrowski, Figs. 3b and 3d) is shown. Synthesis in the former was faster; $m(dAA/dBB) = 9.37 \times 10^{-6}$ $M^{½}s^{-1}$ and $m(dAB) = 2.44 \times 10^{-6}$ $M^{½}s^{-1}$ (Davis, 1998). At ½-order kinetics, there is a steady state when dAA/dBB has a (monomer, or polymer scale) mole fraction ($p(A)$) of 0.937; $p(B)$ is $(1 - p(A))$ and is then 0.063. The S-sector of the 'figure 8' formed by these frequency trajectories contains states with $p(A)$ values less than 0.937. Included in this sector is the symmetry state; $p(A) = p(B) = 0.5$.

FIGURE 2

Species frequencies in the S-sector (Fig. 2) change in a way that elevates dAA/dBB and reduces dAB. As dAA/dBB has the higher formation rate constant, S-sector states are Darwinian, in the sense that they produce positive time variations in mean fitness. N-Sector trajectories (frequencies on right of steady state; p(A) > 0.937), conversely, decrease dAA/dBB and increase dAB. Mean fitness accordingly decreases with time in this sector. In this pattern of change all frequencies on one arm (from steady state to the limit frequency) of each trajectory alter fitness in the same direction with time. The frequency trajectories of a pair of competing species thus exhibit a mirror symmetry about the abcissa, with an intersection at steady state.

DNA synthesis actually departed from ½-order kinetics in these experiments (Davis, 1998). Steady state was shifted to p(A) ≈ 0.60, as a result. This decreased S-sector states and so reduced the range of frequencies that give rise to Darwinian evolution. It is of interest to set limits on the range of steady state frequencies and this issue is taken up in Section 5.

## 4. Equation for rate of change in mean fitness

The rate of change in mean fitness during competitive replication at sub-linear propagation rates is non-decreasing

$$\bar{m}' = S^{y-1}(\widetilde{m^2} - \tilde{m}\bar{m}) \geq 0 \qquad (1)$$

when molecular species have a symmetric frequency distribution. This inequality is extended in Section 5 to all states in the S-sector of the frequency trajectory (Fig. 2). The first term on the rhs of the rate equation

(Davis, 1996a, eqns. 3, A4; Davis, 1996b, eqn. 9) is defined (Davis, 1996a, p. 88) as

$$\tilde{m}^2 \equiv \Sigma_i m_i^2 \wp_i^{y-1} p(i) = \Sigma_i (\nu_i/\bar{\nu}) m_i^2 \wp_i^y \qquad (2)$$

In the notation of Eigen (1971), this term is

$$\Sigma_i k_i^2 x_i^p \qquad (3)$$

apart from inclusion in eqn. (2) of a scale factor, $\nu_i/\bar{\nu}$; $\nu_i$ being the number of 'monomers' in sequence-i and $\bar{\nu}$ is the mean number of 'monomers' per strand. $m_i$ is a formation rate constant; at linear propagation rates it represents a Malthusian fitness parameter. $\wp_i$ is the polymer-scale mole fraction of i-sequence, $S_i/S$. $p(i)$ is their monomer-scale frequency; $p(i) = \nu_i S_i/\bar{\nu} S$ (Davis, 1996a, p. 71), where $S_i$ is the concentration of single strands (replicative form) of i-type oligomers and $S$ is $\Sigma_i S_i$. y refers to the order of the condensation reaction with respect to template concentration. $m_i$ is $k_i$, $\wp_i$ is $x_i$ and y is p, in the notation of Eigen (1971). The product of probabilities, $\wp^{y-1} p(i)$, in eqn. (2) corresponds to

$$\wp^{y-1} p(i) = (S_i/S)^{y-1} (\nu_i S_i/\bar{\nu} S) = (\nu_i/\bar{\nu}) \wp^y \qquad (4)$$

This yields the y-mean square rate constant at monomer scale (eqn. 2) as the first rhs term for time variations in mean formation rate constant (eqn. 1).

The second rhs term in eqn. (1) is a product of the y-mean and arithmetic mean for rate constants; $\tilde{m} = \Sigma i (\nu_i/\bar{\nu}) m_i \wp_i^y$ and $m = \Sigma_i m_i p(i)$. At first order ($y = 1$), the rhs of eqn.(1) equals the statistical variance in formation rate constants, consistent with Fisher's theorem.

Varga and Szathmary (1997, eqn. 8; the rate equation given lacks a time derivative) misinterpret the first rhs term of the rate equation (eqn. 1). Rather than the term defined (Davis, 1996a, p.88; eqn. 2), they state it as (Varga and Szathmary, eqn. 10)

$$\Sigma_i (\nu_i/\bar{\nu}) m_i^2 \wp_i^{2y}$$

allowing for changes in notation and scale. Thus, they give it as the square of the y-mean rate constant (exponent above tilde), rather than the required y-mean of square rate constants (exponent below tilde). Their error could arise from incorrectly composing the probability product, $\wp^{y-1} p(i)$. Instead of $\wp^{2y}$, as they state, the product gives $\wp_i^y$, omitting the scale factor (Davis, 1996a, p.88; eqn. 4).

## 5. Generalized symmetry principle

The rate equation specifying changes in mean fitness under fractional order kinetics (eqn. 1) indicates that time variations in the mean comply with the following inequalities (Davis, 1996a,b),

$$\tilde{m}^2 > \tilde{m}\bar{m} \Rightarrow \bar{m}' > 0 \qquad (5a)$$

$$\tilde{m}^2 < \tilde{m}\bar{m} \Rightarrow \bar{m}' < 0 \qquad (5b)$$

$\tilde{m}^2$, $\tilde{m}$ and $\bar{m}$ are the y-mean of squared fitness, y-mean and arithmetic mean of fitness (Section 4). Time variations in frequency of species-i are then (Davis, 1996a, eqn. A3)

$$m_i \wp_i^{y-1} > \tilde{m} \Rightarrow p(i)' > 0 \qquad (6a)$$

$$m_i \wp_i^{y-1} < \tilde{m} \Rightarrow p(i)' < 0 \qquad (6b)$$

The condition for steady state given by Szathmary and Gladkih (1989) and its revised form (Davis, 1996b, eqn. 3) are

$$\hat{\wp}_i^{y-1} = m_{max}/m_i \qquad \text{(Szathmary-Gladkih)} \qquad (7a)$$

$$\hat{\wp}_i^{y-1} = \Sigma i (\nu_i/\bar{\nu}) m_i \wp_i^{y} / m_i \qquad \text{(revised)} \qquad (7b)$$

The revised steady state condition is scaled to monomer incorporation, consistent with the rate equation for fitness (eqn. 1). At steady state, any two species can be equated;

$$m_i \hat{\wp}_i^{y-1} = m_j \hat{\wp}_j^{y-1} \qquad (8a)$$

$$\left(\frac{m_i}{m_j}\right)^{1/(y-1)} = \left(\frac{\hat{\wp}_j}{\hat{\wp}_i}\right) \qquad (8b)$$

Steady state frequency for a species in pairwise competition is

$$\hat{p}_i = (1 + (m_i/m_j)^{1/(y-1)})^{-1} \tag{9}$$

In a population with s kinds of species, steady state frequency for any particular species can be inferred to be

$$\hat{p}_i = m_i^{-1/(y-1)} \Big/ \sum_{j=1}^{s} (m_j^{-1/(y-1)}) \tag{10}$$

An equivalent expression for steady state frequency was obtained by Varga and Szathmary (1997, p. 1152).

The range of steady state frequencies for two molecular species, i and j, can be deduced from eqn. (8). We consider $m_i > m_j$,

$$\left(\frac{m_i}{m_j}\right)^{-1/(y-1)} = \left(\frac{1 - \hat{p}_i}{\hat{p}_i}\right), \quad \hat{p}_i + \hat{p}_j = 1, \quad 0 < y < 1 \tag{11a}$$

$$\lim_{(m_j/m_i) \to 1} \left(\frac{m_i}{m_j}\right)^{-1/(y-1)} = 1 \Rightarrow \hat{p}_i = \hat{p}_j = 0.5 \tag{11b}$$

$$\lim_{(m_j/m_i) \to 0} \left(\frac{m_i}{m_j}\right)^{-1/(y-1)} = 0 \Rightarrow \hat{p}_i = 1, \hat{p}_j = 0 \tag{11c}$$

Steady state frequency for high and low fitness species thus exist within the limits, $0.5 \leq \hat{p}_i \leq 1$ and $0 \leq \hat{p}_j \leq 0.5$. Both species are equifrequent at steady state ($\hat{p}_i = \hat{p}_j = 0.5$) when their rate constants are equal; $m_i = m_j$. Mean fitness is then invariant ($\bar{m}' = 0$) with changes in frequency.

In a system of s species, the lower limit for the steady state frequency distribution occurs at uniformly

equal fitness

$$\lim_{(m_i/m_j) \to 1} \left( \frac{m_i^{-1/(y-1)}}{\sum_j (m_j^{-1/(y-1)})} \right) = \frac{1}{s} \Rightarrow \hat{p}_i = \frac{1}{s} \quad \forall_{i=1}^{s} \quad (12a)$$

Steady state frequency achieves its highest value when one species, s, has far higher fitness, $m_s$, than all other species in the system

$$\lim_{(m_i/m_s) \to 0} \left( \frac{m_i^{-1/(y-1)}}{\sum_j (m_j^{-1/(y-1)})} \right)_{i \neq s} = 0 \Rightarrow \begin{cases} \hat{p}_i = 0 \quad \forall_{i=1}^{S-1} \\ \hat{p}_s = 1 \end{cases} \quad (12b)$$

An equifrequent species distribution is the lower frequency limit for steady state. It indicates all species have equal fitness. Any system with an equifrequent distribution of species, which differ in fitness will consequently produce positive time variations in mean fitness;

$$\bar{m}(t) < \bar{m}_{ss} \Rightarrow \bar{m}(t)' > 0 \quad \begin{aligned} p(1) &= p(2) = \ldots p(s) \\ m_1 &= m_2 = \ldots m_s \end{aligned} \quad (13)$$

Pairwise competition between species i and j at non-stationary frequencies gives rise to positive time variations in their mean fitness when

$$\bar{m}(t)' = p(i)'(m_i - m_j) > 0 \quad (14)$$

the rate term is positive, $p(i)' > 0$ and $m_i > m_j$; $p(i) + p(j) = 1$. For i and j equifrequent, $p(i)' \geq 0$ for all pairs with unequal fitness. Invariant frequencies ($p(i)' = 0$) mark the lower limit of the steady state range; $m_i = m_j$

(eqn. 11b). Apart from this state, p(i)' will be positive (eqn. 14). Given that $m_i > m_j$, eqn.(11a) requires that $\hat{p}_i$ exceed $\hat{p}_j$. This implies that $\hat{p}_i$ also exceeds $\wp_i$ in any pre-steady state, when i and j are a pair of initially equifrequent, non-neutral species. Time variations in p(i) will be positive, therefore, as the system evolves toward steady state. The two-fold symmetry of frequency trajectories (Fig. 2) requires that species-i frequency, at any point on the arm containing the equifrequent state, will increase with time; $p_{S\text{-sector}}(i\text{-arm})' > 0$. Every p(j) on the j-arm of the S-sector trajectory will conversely decrease with time; $p_{S\text{-sector}}(j\text{-arm})' < 0$. As $m_i > m_j$, it follows that

$$\bar{m}(t) < \bar{m}_{ss} \quad \Rightarrow \quad \bar{m}(t)' > 0 \qquad (S\text{-sector}) \qquad (15a)$$

for all states in the S-sector. The generalized symmetry principle may be stated as: *All states in the frequency trajectory sector encompassing the equifrequent distribution display positive time variations in mean fitness during competitive replication between a pair of species with unequal fitness at sub-linear propagation rates*.

All sub-steady state i-frequencies ($p(i) < \hat{p}(i)$) occur on the upper trajectory arm in the S-sector (Fig. 2) and they have positive time variations; $p(i)' > 0$. Conversely, only above steady state i-frequencies occur on the lower trajectory arm in the N-sector; $p(i) > \hat{p}(i)$. Time variations in p(i) are then negative; $p(i)' < 0$. With $m_i$ larger than $m_j$, pre-steady state mean fitness, $m(t)$, exceeds its steady state value;

$$\bar{m}(t) > \bar{m}_{ss} \Rightarrow \bar{m}(t)' < 0 \quad \text{(N-sector)} \quad (15b)$$

Negative time variations in m are assured, given that p(i) decreases and p(j) increases as they evolve to steady state.

Extension to a general frequency distribution follows on noting that eqn. (14) holds for each pair of species in a system. The frequency of a species-i in the whole system is

$$P(i) = \frac{\nu_i n_i}{\bar{\nu} n} = \phi(i,j) p(i), \quad \phi(i,j) = \frac{\nu_i n_i + \nu_j n_j}{n} \quad (16)$$

$$p(i) = \frac{\nu_i n_i}{\nu_i n_i + \nu_j n_j}$$

Mean fitness then takes the form

$$\bar{m} = \Sigma_{i,j} \phi(i,j)(p(i)(m_i - m_j) + m_j), \quad m_i > m_j, \quad (17)$$
$$j = 1,2,\ldots s/2, \quad i = (s/2+1),(s/2+2)\ldots s$$

Differentiation of each product in the sum with respect to time gives the rate of change in mean fitness,

$$\bar{m}' = \Sigma_{i,j} \phi(i,j)(p(i)'(m_i - m_j)) + $$
$$\Sigma_{i,j} \phi(i,j)'(p(i)(m_i - m_j) + m_j) \quad (18)$$

In a system with symmetrically distributed species, all p(i)' are positive. More generally, p(i)' are positive for every i, when all species frequencies are in the S-sector.

Conversely, p(i)' is negative for all N-sector states. The sign of the first rhs term is, therefore, indicative of the direction of evolution in a heterogeneous system.

The second term (eqn. 18) contains a factor, $(p(i)(m_i - m_j) + m_j)$, which is positive in all distributions. When this factor is equal in all pairs, the second term will vanish. With the second factor a constant, c, the second term is

$$c \sum_{i,j} \phi(i.j)' = c \sum_{j=1}^{s/2} \sum_{i=s/2+1}^{s} (P(i)' + P(j)') = c \sum_{i=1}^{s} P(i)' = 0 \quad (19)$$

In general the second term in eqn. (18) will be small. Time variations in mean fitness for a distribution containing an odd number of species may be obtained using the representation in eqn. (18) by creating one virtual pair with p(j) and $m_j$ set at zero.

## 6. Fitness threshold and survival of the fittest set of species

The steady state frequency of a species is fitness dependent. This dependence is specified by eqn. (10) for a system with s coexisting species. Consonant with this relation, steady state is displaced far to the right in Fig. 2. dAA/dBB fitness exceeded dAB fitness by almost four-fold, during replicase-free synthesis of these DNA hexamers (Sievers and von Kiedrowski, 1994); m(dAA/dBB) and m(dAB) were, respectively, $9.37 \times 10^{-6}$ and $2.44 \times 10^{-6}$ $M^{\frac{1}{2}}s^{-1}$ (Davis, 1998). Their steady state frequencies were estimated as 0.937 and 0.063, assuming ½-order kinetics. With this disparity in steady state frequencies, a small system is likely to exclude, dAB. The Poisson probability, $(1 - e^{-p(dAB)})$, of sampling at least 1 dAB molecule indicates the minimum sample size is nearly 16 DNA molecules. Sub-threshold systems should be homogeneous for dAA/dBB, rather than polymorphic.

From eqn. (10), the frequency threshold imposes the following constraint on frequency,

$$n\wp_i = \left[ \frac{n_i m^{-1/(y-1)}}{\Sigma_i (m_i^{-1/(y-1)})} \right] \not< 1 \, , \, 0 < y < 1 \quad \text{(frequency limit)} \quad (20)$$

n refers to the total number of replicating molecules. For any molecular system there is a cut-off frequency; $\wp^* = 1/n$. Species having a sub-threshold frequency should be excluded from the system. Accordingly, only species with fitness exceeding the lower limit,

$$m^* \not< \left[ \frac{n_i}{\Sigma_i (m_i^{-1/(y-1)})} \right]^{y-1} \quad 0 < y < 1 \quad \text{(fitness limit)} \quad (21)$$

can coexist. This inequality indicates that an increase in fitness or a decrease in the number of DNA molecules, n, within a system elevates the lower limit for species fitness at steady state. Species with sub-threshold fitness experience negative selection and are likely to be excluded from steady state. Survival of the fittest thus applies at sub-linear propagation rates. Only the set of species with above threshold fitness will survive. Survivor fitness determines its abundance at steady state (eqn. 10).

Figure 3 depicts the effect of a variant with high fitness on the steady state frequency distribution attained in a model system that is limited to a constant number (100) of replicator molecules at ½-order kinetics. Initially, species S1, S2, S3 and S4 coexist. Their steady state frequencies being 1, 3, 43 and 53, reflecting fitness (formation rate constants) of 0.06, 0.1, 0.36 and 0.4 $M^{½}s^{-1}$, respectively. Formation of variant, V1, with fitness 0.41 $M^{½}s^{-1}$ destabilizes the initial steady state. This forces the frequency of S1 to sub-threshold levels; p(S1) < 0.01. Above average fitness of V1 and elimination of S1, with below average fitness, increases mean fitness from 0.37 to 0.39 $M^{½}s^{-1}$. A new steady state forms with 2 S2, 28 S3, 34 S4 and 36 V1 molecules. The frequency of S2, S3 and S4 has decreased. In particular, S2 now approaches elimination.

FIGURE 3

According to eqn (20), addition of any species with above threshold fitness will lower the frequency of all species in a pre-existing steady state. This moves species with low fitness closer to the frequency threshold, elevating their risk of exclusion from the system. As species with sub-threshold fitness are excluded from steady state, while any species with above threshold fitness is accepted, it is apparent that threshold fitness represents a non-decreasing quantity in competitive replication at sub-linear propagation rates.

This conclusion may be formalized by noting that the sum of species fitness term in eqn. (21) is initially

$$M(0) = \sum_{i=1}^{S} (m_i^{-1/(y-1)}) = m_1^{-1/(y-1)} + m_2^{-1/(y-1)} + \ldots + m_S^{-1/(y-1)}$$
$$m_1 < m_2 < \ldots < m_s, \quad 0 < y < 1$$

Following addition of a variant, V, and exclusion of species S1, the sum of species fitness term becomes after time, t,

$$M(t) = m_v + \sum_{i=1}^{S} m_i^{-1/(y-1)} = m_v^{-1/(y-1)} + m_2^{-1/(y-1)} + \ldots + m_S^{-1/(y-1)}$$
$$m_1 < m_v < \ldots < m_s, \quad 0 < y < 1$$

Hence, $M(t) > M(0)$.

Changes in threshold fitness, $m^*$, are then positive with time,

$$\delta_t m^* = \left( \frac{n}{\delta_t \Sigma_i (m_i^{-1/(y-1)})} \right)^{y-1} \geq 0 \qquad (22)$$

where $\delta_t$ denotes a time variation, $g(t) - g(0)$. Population size, n, being held constant in this model system.

Increases in the fitness threshold accompany addition of even a low fitness variant. Addition of a variant with fitness of 0.11 $M^{\frac{1}{2}}s^{-1}$ to the steady state between S1, S2, S3 and S4 is depicted in Fig. 3. It reduced mean fitness from 0.37 to 0.36 $M^{\frac{1}{2}}s^{-1}$. In contrast, the variant raises the fitness threshold from 0.055 to 0.056 $M^{\frac{1}{2}}s^{-1}$. Fig. 3 demonstrates that addition of a variant with above average fitness will raise both mean and threshold fitness; the mean increased from 0.37 to 0.39 $M^{\frac{1}{2}}s^{-1}$ and the fitness threshold simultaneously increased from 0.055 to 0.068 $M^{\frac{1}{2}}s^{-1}$.

Equations (20, 22) also imply that the interval until elimination of a species during competitive replication at sub-linear propagation rates is fitness dependent. Since the fitness threshold increases during evolution, species possessing progressively higher fitness will be excluded from the system,

$$t_j \propto m_j^{-1/(\gamma-1)} \qquad (23)$$

is the interval until exclusion of species-j with fitness $m_j$.

This principle is illustrated in Fig. 4. A steady state between species S1, S2 and S3, with fitness 0.015, 0.02 and 0.06 $M^{\frac{1}{2}}s^{-1}$, respectively, is shown to be destabilized by mutation to a series of eight variants with higher fitness over an interval of 11 min. S1, S2 and S3 were pushed to sub-threshold frequencies and excluded after evolution for intervals of 703, 715 and 729 s. The length of each exclusion interval corresponds to the fitness ranking of a species, as required for propagation at sub-linear rates (eqn. 22). To obtain an explicit relation between exclusion interval and replicative fitness, error rates and topography of fitness space would need to be considered.

FIGURE 4

## 7. Discussion

A distinction has been made in this investigation between the attractor of an evolving system and the nearest accessible stationary state. In competitive replication at sub-linear propagation rates, a system evolves toward a polymorphic steady state - the revised Szathmary-Gladkih attractor (Davis, 1996b; eqn. 7b). A distribution of species frequencies consistent with the attractor-specified profile was seen to be subject to modification by the system. Those frequencies that correspond to less than one replicator molecule should be omitted. Model prebiotic systems were found to display a steady state distribution truncated at the frequency threshold (Section 6).

Since the steady state frequency attained by a species is fitness dependent, only species with above threshold fitness can be expected to survive competitive replication at sublinear propagation rates. It is evident, moreover, that exclusion of species with subthreshold fitness combined with admission of all species above the threshold will cause the fitness threshold to increase over time (Section 6). Consequently, propagation any operative error rate will result in selection against species with sub-threshold fitness and an increase in the sum of species fitness.

In replicase-free synthesis of complementary (dAA/dBB) and self-complementary (dAB) DNA hexamers (Sievers and Kiedrowski, 1994) virtually all variants would exceed the fitness threshold. Synthesis was initiated in these experiments in a 1 µl capillary containing around $9.64 \times 10^{13}$ DNA molecules (about 0.16 mM). As dAA/dBB and dAB had formation rate constants of $9.37 \times 10^{-6}$ and $2.44 \times 10^{-6}$ $M^{\frac{1}{2}}s^{-1}$, respectively, threshold fitness (eqn. 21) for their steady state was only $9.86 \times 10^{-13}$ $M^{\frac{1}{2}}s^{-1}$; about seven orders of magnitude less than the fitness of dAA/dBB. In view of this extremely low fitness threshold, essentially all variants would be retained by the system, increasing heterogeneity over time. Calculation reveals that a steady state with almost $7 \times 10^{12}$ species, with fitness comparable to dAA/dBB, would elevate the fitness threshold to $2.5 \times 10^{-6}$ $M^{\frac{1}{2}}s^{-1}$. This threshold is sufficient to exclude dAB. Heterogeneity of this level clearly exceeds the capacity of any hexameric quaternary sequence ($4^6$ = 4096). All possible sequences up to 21 to 22 nucleotides (Davis, 1996a, p. 74) yield heterogeneity of this amount, after discounting one-fourth (m(dAB)/m(dAA/dBB) ≈ ¼) of sequences that have sub-threshold fitness. On reducing a model prebiotic system to around 100 molecules, however, only 7 or 8 species, with fitness equal to dAA/dBB, are sufficient to raise the fitness threshold sufficiently to exclude dAB.

By establishing that the fitness threshold is a non-decreasing parameter in sub-linear molecular evolution a link is made between this process and Darwinian evolution. The stationary state fitness threshold in a Darwinian system can be deduced from the expression for frequency invariance in a Darwinian system,

$$p(i)' = p(i)(m_i - \bar{m}_{ss}) = 0 \Rightarrow m_i = \bar{m}_{ss} \; \forall_i$$

where it is seen to equal the stationary state mean fitness. Increases in the fitness threshold within this system shift it toward a homogeneous state of maximum fitness. In this event, threshold fitness equals both the stationary state mean and maximum fitness: $m^* = \bar{m}_{ss} = m_{max}$. Even with a Darwinian attractor a polymorphic stationary state is accessible, when there exist neutral variants. Since Fisher (1930), positive time variations in the mean of fitness among competing alleles and molecular replicators (Davis, 1978) have had a place at the center of evolution theory.

Positive time variations in mean fitness are a less general feature of sub-linear molecular evolution than increases in the fitness threshold. An increasing mean fitness is limited to the symmetry distribution (Davis, 1996a) and other states in the S-sector of the frequency trajectory (Section 3). Frequency changes in N-sector states produce negative rates of change in the mean (Fig. 2). Whereas, the fitness threshold is non-decreasing in both linear and sub-linear (S- + N-sector) systems. Hence, this principle is a more general than Fisher's fundamental theorem of natural selection, although it is not as general as the force damping principle (Davis, 1996a, 1998).

A stable steady state can form among a set of competitively replicating species that propagate at a fixed fractional order with respect to template concetration (Varga and Szathmary, 1997; Davis, 1998). Addition of a variant with above threshold fitness destabilises the steady state. Elimination of low fitness species was demonstrated to occur in the post-steady state transition (Section 5). A continuous stream of variants, formed by errors during propagation, would prevent evolution from being trapped at the first accessible steady state in a sublinear system. Selection against species with sub-threshold fitness, moreover, would elevate the fitness threshold with time. Its non-Darwinian attractor notwithstanding. molecular evolution at sublinear propagation rates therefore parallels Darwinian evolution in this respect.

Since Fisher (1930), positive time variations in the mean of fitness among competing alleles or self-replicating polynucleotide molecules (Davis, 1978) have been known to characterize Darwinian evolution. Displacement of the system to a more probable kinetic state results (Davis, 1996a) and this is a feature of all states in the S-sector of the frequency trajectory (Sections 3, 5). N-sector states produce negative changes in mean fitness, because the kinetic force is coupled to a second force. Replicase-free synthesis of DNA or RNA oligomers (von Kiedrowski, 1986; Zielinski and Orgel, 1987) at fractional order kinetics couples synthesis and duplex formation. The latter is accelerated by advancement of synthesis and duplex formation then retards synthesis. These inter-relationships account for a parabolic time course during synthesis (von Kiedrowski et al., 1991). The decrease in mean fitness resulting from a high-fitness replicator at above steady state frequency can be attributed to the action of the thermodynamic force driving duplex formation. By preferentially retarding synthesis of sequences with high propagation rates, this force drives the system to a non-optimal kinetic state (high effective activation free energy) for replication. A displacement in this direction is consistent with the action of a negative kinetic force. Stability arises at steady state, when an equilibrium exists between the kinetic force, driving a system to its most probable kinetic state, and the thermodynamic force responsible for strand annealing. As displacement of the frequency distribution ceases at steady state, no net force then acts on the distribution. The transition to steady state can consequently be depicted (Davis, 1998) as damping a scalar force based on generalized rate coefficients for synthesis.

My thanks to Elia Anastasiadis, Macquarie University, for his skilful assistance.

**Figure legends**

Figure 1. Comparison of the rate of change in mean fitness in various symmetrical and non-symmetrical frequency distributions during competitive replication with a ½-order dependence on template concentration. a, b, d and e, Non-symmetrical distributions showing negative and positive rates. c, f, g and h, Distributions with translational and mirror (concave-upright bell, convex-inverted- bell) symmetry show positive rates only. c, combines the frequency distributions a and b. f, combines distributions d and e. Each rate has a scale factor of $x10^{-4}$ $M^{½}s^{-1}$.

Figure 2. Shows dependence of rate of change in frequency and mean fitness on mole fraction of two replicators during pairwise competition at ½-order kinetics. Frequency trajectories for species A and B are mirror images with an intersection at a stable steady state. Formation rate constants (fitness) for species A and B were: m(A) = 9.37x $10^{-6}$ $M^{½}s^{-1}$ and m(B) = 2.44x$10^{-6}$ $M^{½}s^{-1}$. Species A increases in frequency in the S-sector, which includes the symmetry state (P(A) = p(B)). Time variations in mean fitness are positive
in this sector. Species B increases in frequency in the N-sector, resulting in negative rates of change in mean fitness. y, partial order of polymer synthesis with respect to template concentration. Arrows indicate direction of change in species frequency with time.

Figure 3. Shows change in steady state frequency distribution after formation of a variant with high fitness during competitive replication at ½-order kinetics. An initial steady state between species S1, S2, S3 and S4 is destabilized by appearance of a variant, V1. A new steady state forms that includes V1 but excludes S1, whose frequency declined to a sub-threshold level (less than one S1 molecule in system of 100 replicator molecules). Mean fitness (mean formation rate constant) increased from 0.37 to 0.39 $M^{½}s^{-1}$ in this transition. +, mean fitness.

Figure 4. Sequence of species eliminations in a post-steady state transition depends on the fitness ranking of each species undergoing competitive replication at ½-order kinetics. S1, S2 and S3 have fitness of 0.015, 0.02 and 0.06 $M^{½}s^{-1}$, respectively, and were driven to sub-threshold frequencies in a system of 1500 replicator molecules, after intervals of 703, 715 and 729 s. Variant fitness was: V1 = 0.067, V2 = 0.07, V3 = 0.83, V4 = 0.085, V5 = 0.12, V6 = 0.129, V7 = 0.145, V8 = 0.16 $M^{½}s^{-1}$.